\documentclass[a4paper,fleqn]{cas-dc}
\usepackage{caption}
\usepackage{float}
\usepackage{placeins}
\usepackage{cuted}
\usepackage{colortbl}   
\usepackage{amssymb}    

\makeatletter
\newcommand{\tablehere}{\def\@captype{table}\captionsetup{labelfont=bf}}
\newcommand{\figurehere}{\def\@captype{figure}\captionsetup{labelfont=bf}}
\makeatother

\usepackage{tabularx}
\newcolumntype{Y}{{\raggedright\arraybackslash}X}

\usepackage{natbib}
\setcitestyle{numbers,square,sort&compress}
\AtBeginDocument{\let\cite\citep}
 
\begin{document}
\let\WriteBookmarks\relax

\newcolumntype{L}[1]{>{\raggedright\arraybackslash}p{#1}}
\newcolumntype{R}[1]{>{\raggedleft\arraybackslash}p{#1}}
\newcolumntype{C}[1]{>{\centering\arraybackslash}p{#1}}
\renewcommand{\topfraction}{.95}
\renewcommand{\bottomfraction}{.95}
\renewcommand{\textfraction}{.05}
\renewcommand{\floatpagefraction}{.75}
\setcounter{topnumber}{5}
\setcounter{bottomnumber}{5}
\setcounter{totalnumber}{10}

\shorttitle{ROS2MR: Connecting ROS 2 Motion Planning to Mixed-Reality Trajectory Visualization}    


\title [mode = title]{ROS2MR: Connecting ROS 2 Motion Planning to Mixed-Reality Trajectory Visualization }


\author[1]{Santiago Berrezueta-Guzman}[orcid=0000-0001-5559-2056]
\credit{Conceptualization, Writing, Supervision}

\author[1]{Habiba-Loai Khalil}[orcid=0009-0007-3085-5105]
\credit{Methodology, Writing, Data Collection}

\author[1]{Andrei Koshelev}[orcid=0009-0000-3786-4994]
\credit{Methodology, Writing, Data Collection}

\author[1]{Stefan Wagner}[orcid=0000-0002-5256-8429]
\credit{Writing, supervision}

\affiliation[1]{organization={Technical University of Munich}, city={Heilbronn}, country={Germany}}



\begin{abstract}
Mixed reality (MR) allows users to inspect planned robot motion within their physical surroundings, but displaying trajectories from robotics software requires converting their joint configurations and timing into a format the MR application can play back. 
This paper presents \textit{ROS2MR}, an extension of the MR-Robotics LAB platform that converts YAML-serialized ROS~2 joint trajectories, including MoveIt motion plans, into the platform's existing JSON playback format. 
The converter validates the input, maps arm joints by name, and resamples the trajectory at a nominal rate of 30~Hz, using cubic Hermite interpolation when joint velocities are supplied and linear interpolation otherwise, while binary gripper commands are sampled separately using a zero-order hold. 
A Python export helper combines sequences of motion plans and gripper states into a single trajectory and checks positional agreement at segment boundaries. 
The MR application was also adapted for Apple Vision Pro in addition to Meta Quest~3. 
A technical evaluation showed that the converter preserved the normalized duration and endpoint configurations for dense, sparse, and irregularly timed inputs. 
With supplied velocities, the position RMSE was approximately $2.91\times10^{-5}$~degrees for all sampling patterns; without them, sparse input increased it to 0.85~degrees, and close positional agreement did not guarantee accurate velocity profiles. 
A pick-and-place sequence retained its open--closed--open gripper order, although off-grid transitions were shifted to the output sampling grid. The resulting file-based pathway enables inspection of ROS-generated trajectories in MR without a live ROS connection.
\end{abstract}


\maketitle

\section{Introduction}\label{sec:introduction}

In robotics, researchers have used augmented and mixed reality (AR \& MR) to display robot information, show intended movements, and support interaction with robots \cite{walker2023vamhri,chang2024survey}.
Users can walk around a virtual robot placed in their physical surroundings to inspect its planned motion from different viewpoints \cite{suzuki2022arrobotics}.

Robot motion planning relies on tools for communication, planning, and visualization.
The Robot Operating System (ROS) facilitates communication among robotic components and supports software reuse \cite{quigley2009ros}, while ROS 2 introduces architectural changes to meet the requirements of modern robotic systems \cite{macenski2022ros2}.
MoveIt helps users configure and integrate motion planning for different robot models \cite{coleman2014moveit}, and the Open Motion Planning Library (OMPL) provides implementations of sampling-based planning algorithms \cite{sucan2012ompl}.
RViz supports desktop visualization of robot models and spatial data \cite{ros2rviz}.
To display planned motion in MR, the joint configurations and timing information must be transferred into a format that the MR application can use.

Several systems have connected robotics software to immersive interfaces.
ROS Reality connects ROS-enabled robots to Unity-based virtual reality (VR) interfaces for teleoperation \cite{whitney2018rosreality}.

Our previous work introduced \textit{MR-Robotics LAB}\footnote{MR Robotics Lab platform: \url{https://mr-lab.se.cit.tum.de/}}, which converted MATLAB-generated robot trajectories into JSON files for playback in a Unity application on Meta Quest 3 \cite{mr_robotics_lab}.
The system combined a web-based converter with a mixed-reality application in which users could place virtual robots in their physical surroundings and observe their motion.

Extending this converter to ROS 2 trajectories requires handling their timing and, where available, velocity information.
ROS joint trajectories specify joint configurations, the times at which they should be reached, and optionally the joint velocities.
Treating each source point as a single frame at a fixed playback rate would make the duration depend on the number of points rather than on the specified timing.
Ignoring supplied velocities can also change the reconstructed motion between configurations.
Sequences of motion plans also require consistent joint names and ordering, agreement at segment boundaries, and handling of discrete gripper commands.

In this paper, we extend MR-Robotics LAB to accept ROS 2 trajectories, including MoveIt motion plans, while keeping the existing JSON playback format.
The converter validates YAML-serialized joint trajectories, maps supported joints by name, and resamples the arm motion at a nominal rate of 30 Hz.
It uses cubic Hermite interpolation when velocities are supplied and linear interpolation when they are absent, and it samples binary gripper commands separately.
A Python export helper combines sequences of MoveIt plans with gripper commands into a single trajectory, so the resulting files can be played back without a live ROS connection.
We also adapt the MR application for Apple Vision Pro, extending headset support beyond Meta Quest 3.
Therefore, this study addresses the following research questions (RQs):
\pagebreak
\begin{description}
\item[RQ1:] How accurately does the conversion preserve joint configurations and output timing when resampling dense, sparse, and non-uniformly timed ROS trajectories to a nominal rate of 30 Hz?

\item[RQ2:] How does the availability of source joint velocities affect the position and velocity profiles of the converted trajectories?

\item[RQ3:] How does sequence export maintain positional agreement at segment boundaries and combine segment timing, and how accurately are binary gripper states and transition times represented in the converted JSON?
\end{description}

In addition to answering the RQs above, this work presents four main contributions:
\begin{enumerate}
    \item It provides a reusable pipeline connecting exported ROS 2 trajectories to the existing MR-Robotics LAB JSON format and mixed-reality application.
    \item It implements and evaluates timing-aware conversion using cubic Hermite interpolation when joint velocities are supplied and linear position interpolation with velocity estimation when they are absent.
    \item It introduces an export helper for combining motion plans with binary gripper states, including positional boundary checks and explicit handling of gripper-transition intervals.
    \item It adapts the mixed-reality application for Apple Vision Pro, including changes to interaction required for deployment on visionOS.
\end{enumerate}

\section{Related Work}\label{sec:relatedwork}

Connecting robot motion planning to MR visualization draws on several lines of research: the software tools used to plan and display robot motion, immersive interfaces for inspecting and programming robots, and the interpolation methods that determine how timed trajectories are reconstructed between source points.

\subsection{ROS-Based Motion Planning and Visualization}

ROS and ROS 2 provide software infrastructure for integrating robotics components \cite{quigley2009ros,macenski2022ros2}.
Coleman et al.\ examined how MoveIt reduces the configuration effort required to apply motion planning to different robots through standardized robot descriptions, automated configuration, and extensible components \cite{coleman2014moveit}.
The OMPL supports reuse through a collection of sampling-based planning algorithms \cite{sucan2012ompl}.

Visualization tools and external interfaces make ROS data accessible in different environments.
RViz supports desktop visualization of robot models and spatial data \cite{ros2rviz}, while Rosbridge provides applications outside the native ROS environment by exposing ROS interfaces \cite{crick2017rosbridge}.
Zea and Hanebeck introduced iviz, a Unity-based application for visualizing ROS data on mobile devices with native augmented reality support \cite{zea2021iviz}.

\subsection{Immersive Interfaces for Robot Motion Inspection}

Surveys of AR and MR in robotics identify the visualization of robot state, intended actions, and environmental information as recurring interface functions \cite{suzuki2022arrobotics,walker2023vamhri,chang2024survey}.
These applications include supervising robots and inspecting planned motion within the user's physical environment.

Whitney et al.\ introduced ROS Reality, a framework connecting ROS-enabled robots to Unity-based virtual reality interfaces for teleoperation \cite{whitney2018rosreality}.
Rosen et al.\ combined ROS Reality, MoveIt, Unity, and Microsoft HoloLens to visualize intended robot arm motion and allow users to specify end-effector goal poses \cite{rosen2019motionintent}.
In their study of 32 participants, the mixed-reality interface improved collision classification accuracy and reduced task completion time compared with a desktop visualization interface.

Guhl et al.\ combined distributed sampling-based path planning with AR visualization to support the verification of industrial robot trajectories on mobile devices \cite{guhl2018visualizing}.
Along similar lines, Rosen et al.\ used MR to display intended robot arm motion and studied how users identified potential collisions \cite{rosen2019motionintent}.

\subsection{Robot Programming in Augmented and Mixed Reality}

Researchers have also used immersive interfaces to create and modify robot motion.
Perez Quintero et al.\ presented an AR interface supporting trajectory specification, virtual motion previews, robot-parameter visualization, and online reprogramming \cite{quintero2018augmented}.
Neves et al.\ investigated MR applications for visualizing preplanned robot paths and creating or adjusting paths through spatial interaction \cite{neves2018application}.

Ostanin and Klimchik presented an MR system combining geometric path planning, trajectory planning, and simulation \cite{ostanin2018interactive}.
Subsequent work connected HoloLens, ROS, Unity, and robotic manipulators to support interactive task creation and trajectory generation \cite{ostanin2020interaction}.
Ostanin et al.\ also investigated full hand tracking for industrial robot programming using a system based on Unity, ROS, and MoveIt \cite{ostanin2022handtracking}.

Gadre et al.\ developed an MR interface in which users created and edited robot motions by placing waypoints \cite{gadre2019enduser}.
Their study compared MR and desktop interfaces with 20 participants performing two pick-and-place programming tasks.
More participants completed the tasks using the mixed-reality interface, and task completion times were shorter.
The study examined the creation and adaptation of robot programs in mixed reality.
Our work addresses a different part of the workflow: students generate trajectories in their existing robotics environment and then export them for inspection in mixed reality.
This keeps trajectory programming separate from the headset interface.

\subsection{Trajectory Timing and Interpolation}

A trajectory contains more than a sequence of joint positions: its timestamps specify when those positions should be reached.
The ROS 2 joint trajectory controller uses different interpolation methods depending on the available data.
Its spline option uses linear interpolation for positions alone, cubic interpolation for positions and velocities, and quintic interpolation when accelerations are also supplied \cite{roscontroltrajectory}.
With position-only inputs, the interpolated velocity can change abruptly at source points.

This matters when converting a trajectory to the sampling rate required by another application.
The converter must calculate positions between the source points, and the available velocity information affects that calculation.
ROS2MR uses linear or cubic Hermite interpolation according to whether joint velocities are supplied.
These are established methods; our evaluation checks their implementation within the conversion workflow.

\subsection{Positioning the Present Work}

Our previous work introduced MR-Robotics LAB, which converted MATLAB-generated robot trajectories into JSON files for mixed-reality playback in a Unity application on Meta Quest 3 \cite{mr_robotics_lab}.
Users could place virtual robots in their physical environment and inspect their motion without access to a physical robot.

The present work adds a file-based ROS 2 input pathway to this workflow, including joint-name mapping, timing-aware resampling, handling of optional joint velocities, and export of motion-plan sequences with binary gripper commands.
Its technical evaluation examines how these operations affect the joint profiles, timing, and gripper states represented in the converted files.

We also adapt the mixed-reality application for Apple Vision Pro while retaining the shared JSON playback format.
The adapted application supports virtual robot visualization, playback controls, and joint statistics within the user's physical environment. Table \ref{tab:related-comparison} summarizes and highlights the contribution of this development against the analyzed previous work. 

\begin{strip}
\centering
\tablehere
\caption{Comparison of ROS2MR with related robot visualization and mixed-reality systems.}
\label{tab:related-comparison}
\small
\setlength{\tabcolsep}{5pt}
\renewcommand{\arraystretch}{1.15}
\begin{tabular}{|l|c|c|c|c|c|c|}
\hline
\textbf{System} &
\textbf{\shortstack{Trajectory\\vis./playback}} &
\textbf{\shortstack{ROS\\integration}} &
\textbf{\shortstack{In-situ\\AR/MR}} &
\textbf{\shortstack{Motion authoring\\in headset}} &
\textbf{\shortstack{Playback without a\\live ROS connection}} &
\textbf{\shortstack{Timing-aware\\resampling}} \\
\hline
RViz \cite{ros2rviz} & \checkmark & \checkmark & -- & -- & -- & -- \\
iviz \cite{zea2021iviz} & \checkmark & \checkmark & \checkmark & -- & -- & -- \\
ROS Reality \cite{whitney2018rosreality} & -- & \checkmark & -- & -- & -- & -- \\
Rosen et al.\ \cite{rosen2019motionintent} & \checkmark & \checkmark & \checkmark & \checkmark & -- & -- \\
Guhl et al.\ \cite{guhl2018visualizing} & \checkmark & -- & \checkmark & -- & -- & -- \\
Perez Quintero et al.\ \cite{quintero2018augmented} & \checkmark & -- & \checkmark & \checkmark & -- & -- \\
Gadre et al.\ \cite{gadre2019enduser} & \checkmark & \checkmark & \checkmark & \checkmark & -- & -- \\
Ostanin et al.\ \cite{ostanin2020interaction,ostanin2022handtracking} & \checkmark & \checkmark & \checkmark & \checkmark & -- & -- \\
MR-Robotics LAB \cite{mr_robotics_lab} & \checkmark & -- & \checkmark & -- & \checkmark & -- \\
\hline
\rowcolor[gray]{0.9}
\textbf{ROS2MR (this work)} & \checkmark & \checkmark & \checkmark & -- & \checkmark & \checkmark \\
\hline
\multicolumn{7}{@{}l}{\footnotesize\textit{Note:} ``--'' indicates that the feature is not a primary function of the cited system.} \\
\end{tabular}
\end{strip}

\section{Methodology}\label{sec:methodology}

The ROS 2 extension builds on the existing MR-Robotics LAB workflow, adding trajectory export, input validation, timing-aware resampling, and gripper handling before the shared JSON generation and playback stages. A technical evaluation then examines the resulting conversions, and the MR application is adapted for deployment on Apple Vision Pro.

\subsection{System Overview}\label{sec:system-overview}

\begin{figure*}
    \figurehere
    \includegraphics[width=\textwidth]{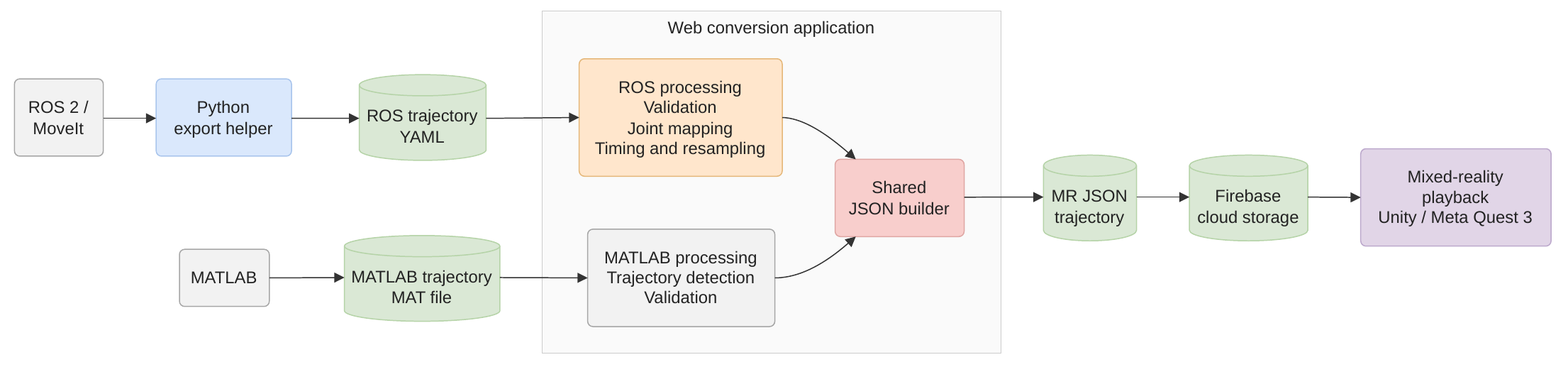}
    \caption{Trajectory conversion workflow in MR-Robotics LAB.
    The ROS 2 pathway adds a Python export helper and timing-aware trajectory processing, while both input pathways share JSON generation, cloud storage, and mixed-reality playback.}
    \label{fig:conversion-workflow}
\end{figure*}

The ROS 2 extension connects trajectory generation to the existing MR-Robotics LAB conversion and playback workflow through three stages: trajectory export, web-based conversion, and mixed-reality playback.
Figure~\ref{fig:conversion-workflow} shows the ROS and MATLAB input pathways and the components they share for JSON generation, cloud storage, and playback.

A Python export helper serializes a ROS joint trajectory or combines a sequence of MoveIt plans into a YAML file, which the user uploads to the web application.
The user selects the corresponding virtual robot model and, when a gripper channel is present, a gripper model.
The converter validates the input, reorders the arm joints by name, and resamples the arm motion at a nominal rate of 30 Hz, while sampling the optional binary gripper channel using a zero-order hold.
A shared JSON builder generates the playback file, which is stored in Firebase and retrieved by the Unity application for mixed-reality playback on Meta Quest 3 or Apple Vision Pro.

\begin{figure*}[t]
    \centering
    \includegraphics[width=\textwidth]{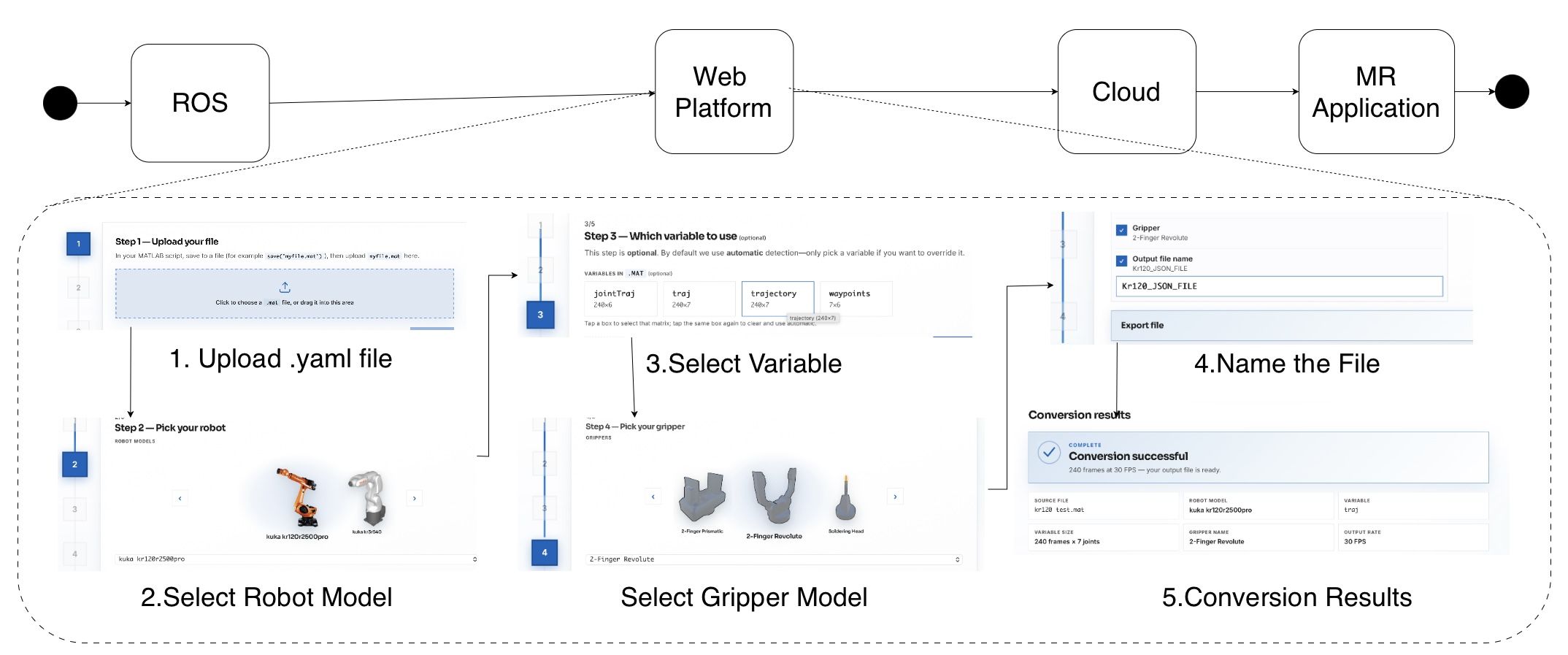}
    \caption{Web interface for trajectory conversion in MR-Robotics LAB \cite{mr_robotics_lab}.}
    \label{fig:web-overview}
\end{figure*}

We chose file-based exchange to connect the existing web and Unity applications without requiring ROS or MATLAB engines to run on the backend.
ROS is used to generate the input trajectory, but playback does not require a live ROS connection.
Figure~\ref{fig:web-overview} illustrates the web-based conversion workflow.

Table~\ref{tab:supported-robots} lists the robot models available in the mixed-reality application; the models used in the technical evaluation are identified separately.
The payload and reach columns describe the corresponding physical robots, based on KUKA product documentation \cite{kukaPortfolio,kukaKR3,kukaKR6,kukaKR120,kukaAgilus,kukaKR560} and the Universal Robots UR5e datasheet \cite{ur5eSpecifications}.
Payload values follow the cited manufacturer listings and include both rated and maximum payload figures.

\begin{table}[t]
    \centering
    \caption{Robot models available in MR-Robotics LAB.}
    \label{tab:supported-robots}
    \small
    \setlength{\tabcolsep}{3pt}
    \begin{tabular}{@{}llrr@{}}
        \toprule
        Model & Class & Payload (kg) & Reach (mm) \\
        \midrule
        \multicolumn{4}{@{}l}{\textbf{KUKA}} \\
        KR3 R540 & Industrial & 3 & 541 \\
        KR6 R900 sixx & Industrial & 6 & 901.5 \\
        KR16 R1610-2 & Industrial & 16 & 1612 \\
        KR20 R1810-2 & Industrial & 20 & 1813 \\
        KR 120 R2500 pro & Industrial & 120 & 2496 \\
        KR150 R2700-2 & Industrial & 150 & 2701 \\
        KR10 R900-2 & Industrial & 11 & 901 \\
        KR50 R2100 & Industrial & 50 & 2101 \\
        KR70 R2100 & Industrial & 70 & 2101 \\
        KR300 R2700-2 & Industrial & 300 & 2701 \\
        KR560 R3100-2 & Industrial & 560 & 3100 \\
        LBR iisy 15 R930 & Collaborative & 15 & 930 \\
        \midrule
        \multicolumn{4}{@{}l}{\textbf{Universal Robots}} \\
        UR5e & Collaborative & 5 & 850 \\
        \bottomrule
    \end{tabular}
\end{table}

\subsection{Trajectory Export}\label{sec:trajectory-export}

For single-trajectory export, the Python helper accepts a ROS joint trajectory message and serializes its joint names, positions, timestamps, and available optional fields without resampling the motion or changing its units.
Any existing gripper channel is retained during serialization, while the sequence interface additionally accepts supported MoveIt containers that expose a joint trajectory.
Interpolation and conversion to the MR playback representation are performed later by the web converter.

For multi-stage tasks, the helper combines an ordered sequence of motion plans paired with binary gripper states.
Each segment must contain six arm joints with consistent joint names and ordering, ensuring that corresponding joint values refer to the same joints throughout the sequence.
To check positional agreement at segment boundaries, the helper compares the initial configuration of each subsequent segment with the preceding segment's final configuration and rejects boundaries where any corresponding joint positions differ by more than $10^{-4}$ rad.
This check enforces positional agreement within the specified tolerance, but does not guarantee velocity continuity between segments.

The helper offsets the timestamps of subsequent segments to construct a combined timeline, omitting duplicate boundary samples where appropriate.
The helper constructs a separate export trajectory so that preparing the visualization does not modify the original motion plans generated by the students' scripts.

\subsection{Input Validation and Joint Mapping}

The converter accepts ROS joint trajectories serialized in YAML. It requires at least two points, position arrays matching the declared joint count, and finite values for the mapped arm joints and any recognized gripper channel. Optional velocity, acceleration, and effort fields are treated as absent when omitted, set to null, or provided as empty arrays; otherwise, they must be arrays matching the declared joint count and contain finite values for the mapped arm joints. Inconsistent coverage across trajectory points is rejected rather than silently selecting a different conversion policy.

Each point must provide an explicit \texttt{time\_from\_start} object containing valid seconds and nanoseconds components. Empty or incomplete timestamp objects, negative durations, and nanoseconds outside the range $[0,10^9)$ are rejected. A zero timestamp is valid, but successive timestamps must be strictly increasing. These checks prevent malformed timing data from being silently normalized into an accepted trajectory.

Six arm joints are identified by name and reordered into the application's joint slots. Supported naming schemes include UR5e joint names, KUKA \texttt{joint\_a1}--\texttt{joint\_a6}, and generic \texttt{joint1}--\texttt{joint6}. Mapping by name allows supported trajectories to be converted independently of their input array order.

An optional gripper channel is identified by the names \texttt{gripper}, \texttt{finger\_joint}, \texttt{tool}, or \texttt{tool0}, with case-insensitive matching. Its input position need not be the seventh array entry. The converter extracts the recognized channel separately from the six reordered arm channels and requires its commands to use the binary convention $0$ for open and $1$ for closed. Arm positions and velocities remain in radians and radians per second during trajectory processing.

\subsection{Timing-Aware Resampling}
\label{sec:trajectory-resampling}

The converter subtracts the initial source timestamp from all trajectory timestamps so that the converted trajectory begins at zero, removing any initial delay before the first source point.

The existing Unity playback application requires trajectories at a nominal rate of 30 Hz. To meet this requirement, the converter generates $N=\max(2,\lfloor 30T+0.5\rfloor+1)$ samples for normalized duration $T$.
The first $N-1$ timestamps follow the grid $t_k=k/30$, for $k=0,\ldots,N-2$, while the final timestamp is set to the endpoint $T$.
The output therefore follows a nominal sampling rate of 30 Hz, with a potentially different final sampling interval.

The converter interprets input points as timed joint configurations and selects the arm interpolation method according to the availability of joint velocities.
When every point includes velocities, it uses cubic Hermite interpolation.
For one joint between consecutive timestamps $t_i$ and $t_{i+1}$, let $q_i$ and $q_{i+1}$ denote the source positions, and $v_i$ and $v_{i+1}$ the corresponding velocities.
With $\Delta t=t_{i+1}-t_i$ and $s=(t-t_i)/\Delta t$, the interpolated position is
\begin{equation}
\begin{split}
q(t) ={}&
(2s^3-3s^2+1)q_i\\
&+(s^3-2s^2+s)\Delta t\,v_i\\
&+(-2s^3+3s^2)q_{i+1}\\
&+(s^3-s^2)\Delta t\,v_{i+1}.
\end{split}
\label{eq:hermite}
\end{equation}

The output velocity is obtained by differentiating this interpolant with respect to time.
The interpolant matches the supplied positions and velocities at both ends of each source interval without additional smoothing.

When velocities are absent throughout the input, positions are interpolated linearly:
\begin{equation}
q(t)=(1-s)q_i+s q_{i+1}.
\label{eq:linear}
\end{equation}

Output velocities are estimated from the resampled positions using central differences at interior samples, a forward difference at the first sample, and a backward difference at the last sample, without additional smoothing.
The two interpolation methods can produce different motion between source points, particularly when those points are widely spaced.
Source points that do not coincide with output timestamps contribute to the interpolation but are not necessarily retained as individual JSON frames.

\begin{figure*}[t]
    \centering
    \includegraphics[width=0.85\textwidth]{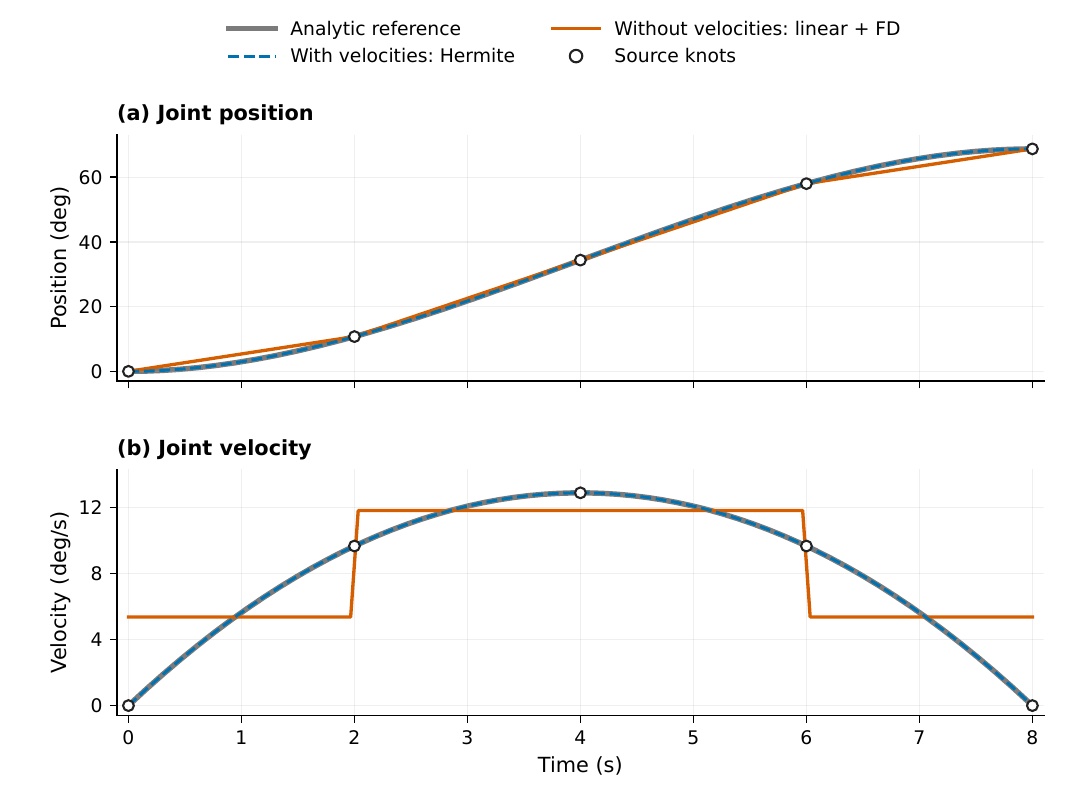}
    \caption{Joint 6 position and velocity for an eight-second synthetic cubic trajectory sampled at $t=\{0,2,4,6,8\}$~s.
    Hermite interpolation with supplied velocities overlaps the reference at this scale; position-only conversion uses linear interpolation and finite differences (FD).
    Circles mark source positions and supplied velocities.}
    \label{fig:resampling-comparison}
\end{figure*}

Figure~\ref{fig:resampling-comparison} illustrates the effect of supplied velocities on the reconstructed position and velocity profiles for a sparsely sampled trajectory.

\subsection{Gripper States and Output Representation}

For sequence export, the helper encodes gripper commands in an additional channel named \texttt{gripper}, using $0$ for open and $1$ for closed. The helper appends this channel to newly constructed trajectory points without modifying the original MoveIt plans. Populated optional arrays receive a zero-valued gripper entry, while empty arrays remain empty.

When the gripper state changes between segments, the helper inserts a point $1/30$~s after the preceding endpoint, repeating the arm configuration and recording the new gripper state.
Any populated velocity, acceleration, and effort arrays are set to zero at the inserted point, and the additional interval is included in the combined timeline.
When velocities are supplied, the helper rejects the transition if any preceding endpoint arm velocity exceeds $10^{-6}$~rad/s in magnitude.
This restriction limits unintended arm motion during the inserted interval, since equal endpoint positions alone do not guarantee a stationary cubic Hermite interpolant.

During conversion, the gripper channel is resampled using a zero-order hold: its command remains unchanged between source transitions rather than being interpolated continuously.
A $10^{-6}$~s comparison tolerance accommodates timestamp rounding at source boundaries. Commands that do not coincide with output sample times are represented on the output sampling grid, so their converted transition times may differ from the source times. A brief state occurring entirely between output samples may not appear in the converted trajectory.

The shared JSON builder preserves the ROS pathway's explicit binary gripper commands directly, including constant-open and constant-closed channels. The MATLAB pathway retains its existing range-based thresholding policy.

The shared builder converts arm positions and velocities to degrees and degrees per second. Each frame represents the gripper separately as a binary state within its \texttt{gripper} object. The arm angle, velocity, and torque arrays contain six entries and exclude the gripper channel. Metadata records gripper availability and the selected gripper model; the declared joint count is seven when the additional gripper channel is present.

When effort values are supplied at every input point, they are mapped to the arm joints and linearly resampled into the output torque fields. Otherwise, the shared JSON builder approximates the torques, using supplied accelerations when available.

\subsection{Technical Evaluation Design}
\label{sec:technical-evaluation}

The technical evaluation examines trajectory export, YAML-to-JSON conversion, and compatibility with the shared playback schema. It addresses the three research questions through controlled numerical comparisons, sequence checks, and regression tests. These checks do not measure robot-controller execution, cloud transfer performance, or playback timing on the headset.

For RQ1, controlled six-joint trajectories provide known reference positions and velocities over an eight-second duration. For joint $j=1,\ldots,6$, the reference position is
\begin{equation}
q_j(t)=0.2j\left[3(t/8)^2-2(t/8)^3\right],
\label{eq:evaluation-reference}
\end{equation}
with positions in radians and reference velocities obtained analytically. Three input sampling patterns are considered: 401 points at 50 Hz, five points at $\{0,2,4,6,8\}$ s, and seven irregularly spaced points at $\{0,0.17,1.21,2.77,5.03,6.91,8\}$ s. Each pattern is evaluated with and without supplied velocities. Converted positions and velocities are compared with the reference at the exported frame timestamps. Additional timing checks examine frame count, normalized duration, endpoint retention, and durations that do not align with the 30 Hz sampling grid. The cubic reference directly tests Hermite reconstruction; it does not establish accuracy for arbitrary motion profiles.

For RQ2, two stored UR5e trajectory fixtures contain identical positions and timestamps, with analytic velocities included in one fixture and omitted from the other. Both contain 401 points over eight seconds and are generated from a clamped cubic spline through eight joint-space waypoints. The converted arm profiles are compared with the generating spline at the nominal output sample times, before timestamp rounding. These trajectories were generated from predefined joint-space waypoints rather than by MoveIt.

Position and velocity errors are summarized using root mean square error (RMSE) and maximum absolute error across the six arm joints. For either quantity, let $e_{k,j}$ denote the difference between the converted and reference values. The pooled RMSE is
\begin{equation}
\mathrm{RMSE}=\sqrt{\frac{1}{6N}\sum_{k=1}^{N}\sum_{j=1}^{6}e_{k,j}^{2}},
\label{eq:evaluation-rmse}
\end{equation}
where $N$ is the output frame count. Position errors are reported in degrees and velocity errors in degrees per second, matching the JSON representation. Both comparisons use serialized output positions and velocities and therefore include their rounding.
The controlled cubic evaluation uses the serialized frame timestamps, whereas the UR5e spline comparison uses the nominal sample times before timestamp rounding.

For RQ3, exporter checks examine segment-boundary positions, timestamp offsets, duplicate-boundary handling, inserted gripper-transition points, and preservation of the original input plans. 
The stored UR5e pick-and-place sequence is converted from YAML to JSON to check its open--closed--open state order, transition timestamps, and gripper metadata. 
Additional cases examine transitions outside the sampling grid, recognized gripper aliases, invalid gripper values, and constant open or closed channels.

The evaluation also runs the repository's 31 JavaScript regression tests and 19 Python exporter tests. The JavaScript tests exercise parsing, joint mapping, resampling, optional-field validation, timestamp validation, and JSON construction, including preservation of constant and changing binary gripper commands. 
The Python tests use minimal ROS message and serialization stubs to check exporter validation, sequence construction, input-plan preservation, and rejection of inserted gripper transitions when any supplied preceding endpoint arm velocity exceeds $10^{-6}$~rad/s in magnitude.

\subsection{Apple Vision Pro Deployment}

Deploying the application on Apple Vision Pro required changes to rendering, dependencies, and interaction. The application uses Unity’s PolySpatial integration with RealityKit for mixed-reality rendering. The deployed configuration imposed restrictions on custom shaders, text rendering, and skinned meshes. Because the Firebase Unity SDK used in this implementation did not provide the required visionOS support, authentication and storage access were reimplemented using Firebase’s REST APIs. The deployed application did not provide scene bounding boxes or access to camera images. Interaction was adapted to gaze-directed pinch input, without direct access to continuous gaze data.
The interaction rig, robot selection, and hand-anchored menu were therefore redesigned for this platform; the resulting interface is shown in Figure~\ref{fig:vision-pro-playback}.

\begin{strip}
    \centering
    \figurehere
    \setlength{\tabcolsep}{0pt}
    \begin{tabular}{cc}
        \includegraphics[height=6.1cm]{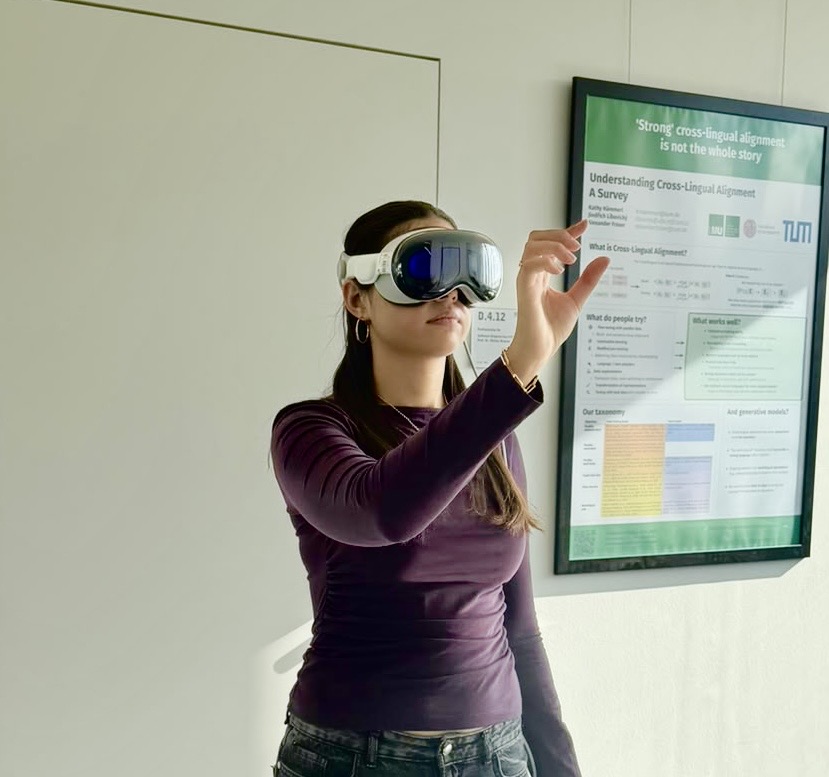} &
        \includegraphics[height=6.1cm]{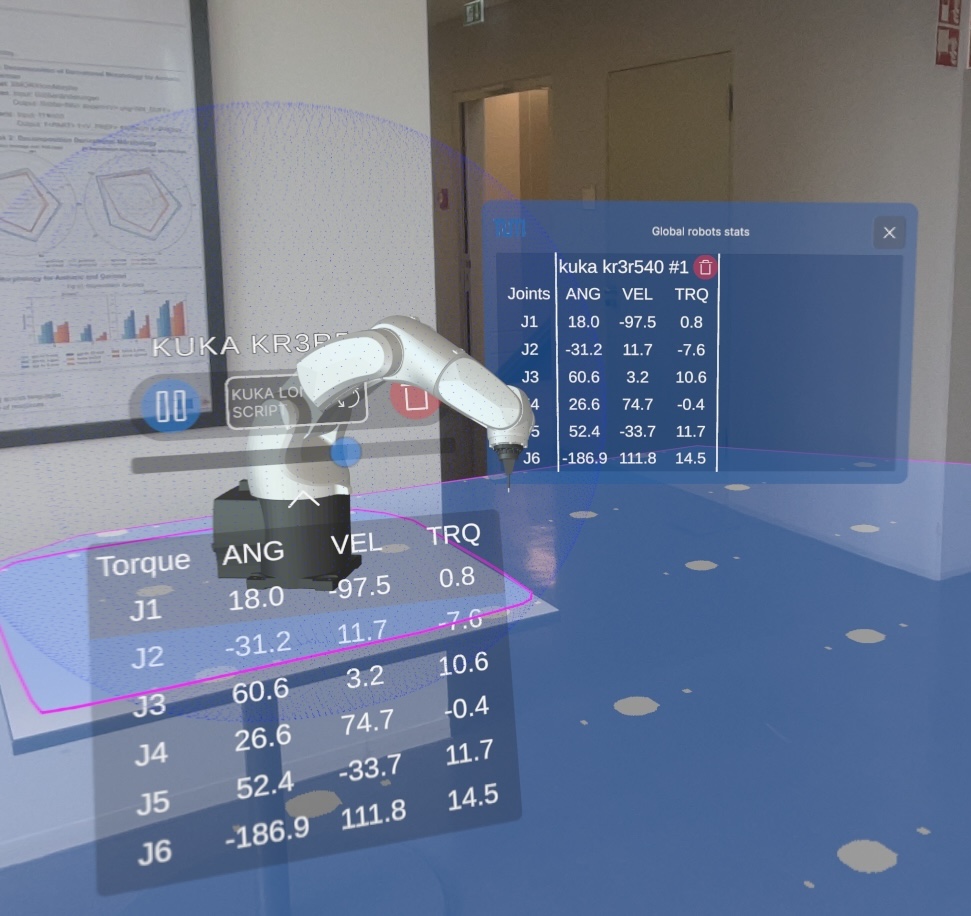} \\[1pt]
        {\small (a) User interaction} & {\small (b) Application view} \\
    \end{tabular}
    \caption{MR-Robotics LAB on Apple Vision Pro.
    (a) A user interacting with the virtual robot while wearing the headset.
    (b) Application view showing a virtual KUKA KR3 R540 robot, playback controls, joint statistics, and spatial overlays.}
    \label{fig:vision-pro-playback}
\end{strip}

\pagebreak
\pagebreak
\section{Results}\label{sec:results}

The evaluation results are organized by research question: conversion accuracy and output timing across sampling patterns (RQ1), the effect of supplied velocities on reconstructed profiles (RQ2), and sequence export with gripper transitions (RQ3), followed by the regression tests covering input validation and joint mapping.

\subsection{Position Accuracy and Timing}
\label{sec:results-position-timing}

All six controlled test cases produced 241 frames over eight seconds, regardless of input density or whether velocities were supplied. Table~\ref{tab:position-errors} reports the position errors relative to the analytic reference across all six arm joints.

\par\medskip
\noindent\begin{minipage}{\linewidth}
\centering
\tablehere
\caption{Position errors for the controlled cubic trajectories. Errors are reported in degrees; supplied velocities are indicated by Yes or No.}
\label{tab:position-errors}
\small
\setlength{\tabcolsep}{4pt}
\begin{tabular}{@{}llrr@{}}
\toprule
Input & Velocities & RMSE & Maximum \\
\midrule
Dense & Yes & $2.91\times10^{-5}$ & $5.33\times10^{-5}$ \\
Sparse & Yes & $2.91\times10^{-5}$ & $5.33\times10^{-5}$ \\
Irregular & Yes & $2.91\times10^{-5}$ & $5.33\times10^{-5}$ \\
Dense & No & $9.20\times10^{-5}$ & $3.29\times10^{-4}$ \\
Sparse & No & $0.8536$ & $2.4243$ \\
Irregular & No & $0.4293$ & $1.4111$ \\
\bottomrule
\end{tabular}
\end{minipage}
\par\medskip

With supplied velocities, cubic Hermite interpolation reconstructed the reference with a position RMSE of approximately $2.91\times10^{-5}$ degrees for every sampling pattern. 
The remaining error reflects output rounding and evaluation at rounded timestamps. 
This small error is expected because Hermite interpolation can reconstruct the chosen cubic reference from its endpoint positions and velocities.

Without supplied velocities, accuracy depended on the spacing of the source points. Dense input produced a position RMSE of $9.20\times10^{-5}$ degrees, whereas sparse and irregular input produced RMSE values of $0.8536$ and $0.4293$ degrees, respectively. 
The sparse case had the largest maximum error, at $2.4243$ degrees. 
The larger errors occurred between source points, where positions and timestamps alone were insufficient to recover the reference motion.

The converter retained the normalized endpoint timestamp, although the final sampling interval was not always $1/30$~s.
A $1.01$~s trajectory produced 31 frames with a final interval of approximately $0.04333$~s, while a $1.02$~s trajectory produced 32 frames with a final interval of $0.02$~s.
Shifting every source timestamp forward by two seconds produced the same output as the zero-offset input, confirming that the converter removes the initial delay.

For RQ1, the converter preserved the normalized duration and endpoint configurations to serialization precision in the tested cases.
Accuracy between source points depended on their spacing and whether velocities were supplied.
These timing results describe the exported JSON, not measured headset playback.

\subsection{Effect of Supplied Velocities}
\label{sec:results-velocities}

Both UR5e test trajectories produced 241 frames over eight seconds, but their converted position and velocity profiles differed despite having the same input positions and timestamps.
With supplied velocities, the converter used cubic Hermite interpolation; without them, it interpolated positions linearly and estimated velocities using finite differences.
Table~\ref{tab:velocity-comparison} reports the errors relative to the cubic spline used to generate the test trajectories.

\par\medskip
\noindent\begin{minipage}{\linewidth}
\centering
\tablehere
\caption{Errors for the paired UR5e spline test trajectories. Position errors are in degrees and velocity errors in degrees per second.}
\label{tab:velocity-comparison}
\small
\setlength{\tabcolsep}{4pt}
\begin{tabular}{@{}llrr@{}}
\toprule
Quantity & Velocities & RMSE & Maximum \\
\midrule
Position & Yes & $2.84\times10^{-5}$ & $5.00\times10^{-5}$ \\
Position & No & $7.06\times10^{-4}$ & $0.002566$ \\
Velocity & Yes & $2.58\times10^{-5}$ & $5.00\times10^{-5}$ \\
Velocity & No & $0.05249$ & $1.0682$ \\
\bottomrule
\end{tabular}
\end{minipage}
\par\medskip

With supplied velocities, position and velocity errors were close to the output rounding scale.
Without them, position errors remained small for this dense trajectory, but velocity RMSE reached $0.05249$~degrees per second, with a maximum error of $1.0682$~degrees per second.
Therefore, matching the joint positions closely did not guarantee an equally close match in velocity.

The controlled cubic cases showed the same dependence on the availability of source velocities. With supplied velocities, velocity RMSE was approximately $2.81\times10^{-5}$ degrees per second for all three sampling patterns. Without supplied velocities, it increased from $0.006969$ degrees per second for dense input to $1.3620$ for sparse input and $0.8683$ for irregular input.

For RQ2, supplying velocities improved reconstruction of the tested reference motion, particularly when source points were widely spaced.
Without velocities, the converter could reproduce positions closely while producing larger velocity errors. Both methods retained the same output duration.

\subsection{Sequence Export and Gripper Transitions}
\label{sec:results-sequences}

The exporter preserved the original input plans, offset the timestamps of later segments, and omitted duplicate boundary samples where appropriate.
It rejected segment boundaries where corresponding joint positions differed by more than $10^{-4}$~rad.
This enforces positional agreement within the tolerance, but does not guarantee velocity continuity between segments.

The stored UR5e pick-and-place YAML contained 403 points across three stages, with the gripper open, closed, and then open again.
Conversion produced 243 JSON frames over approximately $8.066667$~s, rounded to $8.067$~s in the duration metadata.
The two inserted gripper-transition intervals added $2/30$~s to the eight seconds of arm motion.

The resulting JSON retained the open--closed--open state order. The first closed frame occurred at $2.233333$ s, and the subsequent return to open occurred at $5.766667$ s. 
These timestamps agree, within serialization rounding, with the transitions encoded in the YAML. The output metadata declared gripper availability and a joint count of seven, while the arm angle, velocity, and torque arrays retained six entries.

A separate test showed how output sampling affects gripper transition times.
A source command at $0.51$~s first appeared in the JSON at $0.533333$~s, a delay of approximately $0.02333$~s.
The zero-order hold kept the commands binary, but their transition times remained limited by the output sampling grid.

The converter recognized all four supported gripper aliases, rejected non-numeric and non-binary commands, and preserved constant-open and constant-closed channels.
These ROS commands were handled directly rather than through the range-based thresholding used for MATLAB inputs.
The exporter rejected inserted gripper transitions when any supplied preceding endpoint arm velocity exceeded $10^{-6}$~rad/s in magnitude, while accepting transitions at rest.

For RQ3, the exporter enforced positional agreement at segment boundaries, and conversion preserved the tested sequence's combined timeline and open-closed-open gripper order.
Constant gripper states were also preserved, while the off-grid test showed that transition times can be delayed by output sampling.
These results do not establish velocity continuity across every segment boundary or physical executability.

\subsection{Regression Tests and Input Validation}

All 31 JavaScript regression tests and 19 Python exporter tests passed.
The JavaScript tests covered parsing, resampling, joint-name mapping, optional fields, and binary gripper output.
The Python tests checked exporter validation, sequence construction, and preservation of the original input plans using minimal ROS message and serialization stubs.

Reordering input joints under the UR5e, KUKA, and generic naming schemes produced the same converted arm frames as the corresponding ordered inputs.
The converter rejected inconsistent optional-field coverage, malformed or incorrectly sized arrays, non-finite arm values, unsupported joint names, and duplicate or decreasing timestamps.
It also rejected empty or incomplete timestamp objects, negative trajectory durations, and out-of-range nanosecond components, while accepting an explicit zero timestamp.

\section{Discussion}\label{sec:discussion}

The results show what the conversion preserves and where its behavior depends on the input data, the sampling policy, and the scope of the evaluation. These findings bear on how converted trajectories should be interpreted in mixed reality, followed by the limitations of the present work and directions for future development.

\subsection{Timing-Aware Trajectory Conversion}
\label{sec:discussion-timing}

The results demonstrate why ROS trajectory points must be interpreted as timed configurations rather than individual playback frames. 
Dense, sparse, and irregular inputs all produced the same frame count and normalized duration for the eight-second reference trajectory. 
This separates the output timeline from the number of source points and allows trajectories from different sampling patterns to enter the same playback format.

Preserving the normalized endpoint timestamp does not, however, make every output interval identical. The converter uses a 30 Hz grid for intermediate samples and retains the exact endpoint, allowing the final interval to differ from $1/30$ s. 
The initial source delay is also removed. These choices define the visualization timeline rather than preserving every aspect of the original ROS schedule.
Whether the MR application reproduces the exported timing depends on its playback scheduler, which was outside the present technical evaluation.

\subsection{Velocity Availability and Reconstruction Accuracy}
\label{sec:discussion-velocities}

The position and velocity comparisons show that source sampling density and velocity availability jointly affect reconstruction. 
Supplied velocities allow Hermite interpolation to incorporate both configurations and derivatives at source boundaries.
Without this information, linear interpolation provides a simpler reconstruction, but cannot recover an unknown motion profile between widely separated points.

The errors close to the output rounding scale for the controlled cubic reference are consistent with the mathematical properties of Hermite interpolation. 
They should not be interpreted as evidence of equivalent accuracy for arbitrary planned trajectories. The two UR5e test trajectories provide an additional comparison, but they were generated from predefined joint-space waypoints rather than by MoveIt.

The larger velocity errors in the position-only cases also show that close agreement in joint positions does not establish close agreement in derivatives. 
This distinction matters when users inspect velocity profiles or use derived quantities to understand motion.

\subsection{Sequence Composition and Discrete Gripper Commands}

The sequence interface prepares multi-stage tasks for visualization without modifying the original motion plans. 
Its positional boundary check rejects configurations that differ beyond the specified tolerance, while timestamp offsets and inserted gripper-transition points produce a combined timeline. When velocities are supplied, the exporter rejects an inserted transition if any preceding endpoint arm velocity exceeds $10^{-6}$~rad/s in magnitude.
This restriction limits unintended arm movement during the repeated-position hold under cubic Hermite interpolation.
It does not establish velocity continuity across every segment boundary or physical executability of the combined sequence.

Sampling gripper commands separately keeps their values binary and avoids interpolating between open and closed.
Retaining a separate range-based policy for MATLAB inputs allows the shared output schema to accommodate different source conventions without changing the meaning of explicit ROS commands.

Temporal discretization remains a limitation. Transitions outside the output sampling grid are represented at sampled times, and a brief state occurring entirely between samples may be omitted. 
The sequence results therefore demonstrate preservation of the tested command order and constant states, while transition-time accuracy remains constrained by the sampling policy.

\subsection{Compatibility with the Existing Platform}
\label{sec:discussion-platform}

Using the shared JSON builder allows ROS and MATLAB inputs to use the same output schema and existing MR playback application. 
The extension adds a trajectory-authoring pathway without requiring a live ROS connection during visualization.
Joint-name mapping also reduces dependence on the order in which supported arm joints appear in the input arrays.

The regression tests complement the numerical evaluation by checking input validation and discrete conversion behavior. 
They verify rejection of the exercised malformed inputs, preservation of constant and changing binary gripper commands, and enforcement of the inserted-hold velocity constraint.
They do not establish correctness for every possible input or complete end-to-end behavior for every robot model available in the platform.

Both input routes produce files in the format expected by the Unity application.
Sharing this format does not establish equivalence between virtual playback and robot-controller execution or validate the physical accuracy of torque approximations.

\subsection{Limitations}

The evaluation used controlled reference trajectories, stored UR5e test trajectories, and automated tests of the exporter and converter.
It did not include a representative set of MoveIt-generated plans, physical robot execution, headset timing measurements, or a user study.
Joint-name mapping was tested for the supported naming schemes, but the complete export, conversion, and playback workflow was not tested for every robot model available in the application.
The Python exporter tests used simplified ROS message and serialization substitutes rather than a running ROS installation.

The reported errors depend on the trajectories chosen for testing and include output rounding.
Hermite interpolation can reconstruct the controlled cubic reference exactly before rounding, so the small errors in this test should not be expected for all trajectories.
The converter also does not use supplied accelerations to select a higher-order interpolation method and may therefore reconstruct motion differently from a robot controller.

The converter removes the delay before the first source point and adjusts the final sampling interval to retain the normalized endpoint timestamp.
We did not inspect or measure how the Unity application schedules playback, so the timing results describe the exported JSON rather than the motion displayed on the headset.
Gripper commands are sampled at output timestamps, which can shift transition times or omit brief states between samples.

The sequence exporter checks that joint positions agree at segment boundaries within a tolerance, but does not guarantee continuous velocities or a sequence suitable for physical execution.
Its velocity check limits unintended arm movement during inserted gripper holds, but does not validate the dynamics of the combined motion.
Supplied effort values and approximated torques were not checked against physical robot measurements.
The converted files are therefore intended for visualization and should not be treated as equivalent to robot-controller execution.

\subsection{Future Work}

Future testing should include non-cubic reference trajectories and motion plans generated with ROS 2 and MoveIt.
Testing different point densities and timestamp patterns would help determine how well the converter handles a wider range of inputs.
The complete workflow, from trajectory export to retrieval and playback in mixed reality, should also be tested across the supported robot models.

Headset measurements are needed to check whether playback follows the timestamps in the exported JSON, particularly at the final sampling interval and during gripper transitions.
Gripper commands could be stored as separate timed events to avoid losing brief states between trajectory samples.
Another possible extension is to use supplied accelerations for higher-order interpolation, while checking that the output remains compatible with the Unity application.

A user study could examine how students use the ROS extension to inspect planned motion and identify unexpected robot behavior.
It could also test whether displaying position and velocity profiles helps students understand how their trajectories are represented in mixed reality.

\section{Conclusion}

This work presented a ROS 2 extension to MR-Robotics LAB that converts YAML-serialized joint trajectories into the platform's existing JSON playback format. 
The extension combines joint-name mapping, timing-aware resampling, explicit handling of optional velocities, and sequence export with binary gripper commands.
It connects ROS-based trajectory authoring to the established mixed-reality workflow without requiring a live ROS connection during visualization.
The application was also adapted for Apple Vision Pro, including changes to interaction, authentication, and storage access while retaining the shared JSON playback format.

The technical evaluation showed that the converter retained normalized duration and endpoint configurations in the tested cases, while reconstruction accuracy depended on source sampling density and velocity availability. 
Supplied velocities enabled accurate reconstruction of the selected reference profiles; position-only inputs produced larger derivative errors, particularly with sparse sampling. The tested pick-and-place sequence retained its open-closed-open gripper order, and additional regression cases verified preservation of constant binary states. Gripper transition times remained subject to output sampling.

All 31 JavaScript regression tests and 19 Python exporter tests passed, supporting the exercised validation, conversion, and sequence-export behaviors. 
The resulting pathway provides a reusable connection between ROS trajectory authoring and mixed-reality visualization. 
Further testing is needed to assess the complete ROS workflow and determine whether headset playback adheres to the exported timestamps.


\section*{Acknowledgments}

This research was financially supported by the TUM Campus Heilbronn \textit{Incentive Fund 2025} of the Technical University of Munich, TUM Campus Heilbronn. We gratefully acknowledge their support, which provided the essential resources and opportunities to conduct this study.


\bibliographystyle{elsarticle-num}   
\bibliography{00Paper}

\end{document}